\documentclass{aa}  

\newcommand{\ggt}{GG\,Tau}
\newcommand{\uxt}{UX\,Tau}
\usepackage{xcolor}
\usepackage{graphicx}
\usepackage{txfonts}
\begin{document}

   \title{Full orbital solutions in pre-main sequence high-order multiple systems: \ggt\,Ab and \uxt\,B}

   \author{Gaspard Duch\^ene
          \inst{1,2}
          \and
          Jean-Baptiste LeBouquin\inst{1}
          \and
          Fran\c{c}ois M\'enard\inst{1}
          \and
          Nicol\'as Cuello\inst{1}
          \and
          Claudia Toci\inst{3}
          \and
          Maud Langlois\inst{4}
          }

   \institute{Univ. Grenoble Alpes, CNRS, IPAG, 38000 Grenoble, France.\\
              \email{gaspard.duchene@univ-grenoble-alpes.fr}
         \and
             Department of Astronomy, University of California, Berkeley CA, 94720, USA.
        \and
        European Southern Observatory (ESO), Karl-Schwarzschild-Strasse 2, 
              85748 Garching bei Munchen, Germany.
        \and
        CRAL, UMR 5574, CNRS, Université de Lyon, ENS, 9 avenue Charles André, 69561 Saint Genis Laval Cedex, France.
             }

   \date{Received 28/11/2023; accepted 01/04/2024}

% \abstract{}{}{}{}{} 
% 5 {} token are mandatory
 
  \abstract
   {High-order multiple (triple and beyond) systems are relatively common. Their interaction with circumstellar and circumbinary material can have a large impact on the formation and evolution of planetary systems and depends on their orbital properties.}
   {\ggt\ and \uxt\ are two pre-main sequence high-order multiple systems in which the tightest pair has a projected separation of $\approx5$--20\,au. Characterizing precisely their orbits is crucial to establish their long-term stability, to predict the dynamics and evolution of circumstellar matter, and to evaluate the potential for planet formation in such systems.}
   {We combine existing astrometric measurements with previously unpublished high-resolution observations of the \ggt\,Ab and \uxt\,B pairs and perform Keplerian orbital fits.}
   {For \ggt\,Ab the data presented here represent the first detection of orbital motion. For both systems they yield dramatic increases in orbital coverage ($\gtrsim60\%$ and $\approx100\%$ for \uxt\,B and \ggt\,Ab, for orbital periods of $\approx32$ and $\approx8$\,yr, respectively) and allow us to obtain well-constrained orbital fits, including dynamical masses with $\lesssim10\%$ and $\lesssim7\%$ random and systematic uncertainties. We find that both \ggt\,A and \uxt\,A--B likely form stable hierarchical systems, although one possible deprojection solution for \ggt\ is strongly misaligned and could experience von Zeipel-Lidov-Kozai oscillations. We further find that the \uxt\,B orbit is much more eccentric than the \ggt\,Ab one, possibly explaining the lack of circumstellar material in the former.}
   {The newly-determined orbits revive the question of the dynamical fate of gas and dust in these two hierarchical systems and should spur new dedicated simulations to assess the long-term evolution of the systems and the dynamical perturbations imposed by the close binaries they host.}

   \maketitle
%
%-------------------------------------------------------------------
\section{Introduction}

Stellar multiplicity is ubiquitous in the Galaxy, and even more so among pre-main sequence stars \citep{duchene13, offner23}. Since the latter also host planet-forming disks, the influence of stellar companions on planet formation must be studied theoretically, numerically, and empirically. For instance, a long-established prediction is that the orbit of binary stars should severely truncate disks, leading to small circumstellar disks and/or large cavities interior to circumbinary disks \citep[e.g.,][]{artymowicz94}. A direct implication of the former is that circumstellar disks in close binaries should be rare, since small disks have a short (viscous) dissipation timescale, and compact in size. This has been extensively borne out by observations of young binary systems \citep[e.g.,][]{harris12, kraus12, akeson19, manara19}. Similarly, cavities carved out by central binaries have been observed in a number of systems \citep[e.g.,][]{dutrey94, monnier19, ragusa21}. Dynamical perturbations of disks by stellar companions have also been documented in multiple cases \citep[e.g.,][]{kurtovic18, cuello23}.

Among multiple systems, triple and higher-order systems are of particular interest. First of all, such high-order systems represent a sizable fraction of all systems: among field solar-type and low-mass stars, a quarter to a third of all non-single systems contain three or more stars \citep{hirsch21, reyle21}. Furthermore, the growing population of known exoplanets in triple and quadruple systems emphasizes the importance of understanding planet formation in this context \citep{cuntz22}. High-order systems are generally hierarchical in nature because these are inherently more stable \citep[][and references therein]{tokovinin21}. Depending on the relative orientation of the inner and outer orbital planes, they can experience von Zeipel-Lidov-Kozai oscillations, through which orbital eccentricity and the relative alignment angle can vary over broad ranges on secular timescales \citep{vonzeipel10, lidov62, kozai62}. Most importantly, their dynamical influence on surrounding gas and dust is necessarily more complex than that of a binary system, potentially leading to different disk truncation thresholds, complex variability of accretion onto the individual stars, and multiple precession timescales and axes \citep[e.g.,][]{martin22, ceppi22, ceppi23}. GW\,Ori is a well-studied example of a triple-lined spectroscopic system that contains a circumtriple disk whose dynamics is  strongly affected by its orbital configuration \citep{czekala17, bi20, kraus20}. In this study, we focus on wider, visual binaries, which also offer stable locations for individual, circumstellar disks.

\ggt\ and \uxt\ are two hierarchical high-order systems in the L1551 cloud in the Taurus star-forming region. Since they both host multiple disk components, they are ideal test-beds to assess the dynamical effects of multiple systems on their surrounding material. \ggt\ is composed of a hierarchical triple system, \ggt\,A, separated by 10\arcsec\, from a binary system, \ggt\,B \citep{leinert93}. The close binarity of \ggt\,Ab was discovered by \cite{difolco14} and its projected separation of $\approx4$\,au suggested an orbital period of a few years. A massive gas and dust disk surrounds \ggt\,A \citep{guilloteau99, phuong20} and several streamers of material connect it to small circumstellar disks associated with both \ggt\,Aa and Ab \citep{roddier96, silber00, beck12, keppler20}. While it is expected that the orbital motion of the \ggt\,Aa--Ab pair is responsible for carving the cavity inside of the massive disk, there remains some tension between the cavity size and binary orbit, and it is likely that the orbit and disk are not in a coplanar configuration \citep{beust06, kohler11, nelson16, aly18, keppler20, toci24}.

\uxt\ is a quadruple system \citep{reipurth93}, with the close binarity of \uxt\,B first reported by \cite{duchene99}. Subsequent monitoring revealed significant orbital motion, although it remained insufficient for precise orbital fitting \citep{schaefer14}. While \uxt\,B does not host circumstellar matter, \uxt\,A harbors a well-studied transitional disk \citep[e.g.,][]{andrews11, akeson19}. \uxt\,C represents a more complex case: despite a lack of near-infrared excess \citep{white01, mccabe06}, it was found to be associated with both gas and dust \citep{zapata20, menard20}. Most strikingly, these studies identified tidal trails extending from the \uxt\,A disk that suggest a direct dynamical interaction between the two components. Considering hydrodynamical simulations of different configurations, \cite{menard20} concluded that the pair recently underwent an inclined prograde flyby, during which \uxt\,C may have captured material from the \uxt\,A disk. 

In this paper, we present new high-resolution observations of the close \ggt\,Ab and \uxt\,B subsystems (Section\,\ref{sec:obs}), revealing (nearly) full orbital coverage and enabling precise orbital fits (Section\,\ref{subsec:orbits}). From these, we derive dynamical masses for the systems (Section\,\ref{subsec:mdyn}) and constrain their geometry relative to the other components of the systems and to the various disks they host (Section\,\ref{subsec:architecture}). In Section\,\ref{sec:discuss}, we discuss the implications of our findings on the stellar properties, the survival of compact circumstellar disks in close binaries, and the overall stability and dynamics of the systems.

%--------------------------------------------------------------------
\section{Observations and detection of the companions}
\label{sec:obs}

For both systems, our analysis combines published astrometry with new analysis of archival data. In this section, we describe how we processed the latter to determine the relative astrometry at new epochs.

% - - - - - - - - - - - - - - - - - - - - - - - - - - - - - - - - - -
\subsection{\ggt\,Ab}
\label{subsec:obs_gg}

\begin{figure*}
   \centering
   \includegraphics[width=0.75\textwidth]{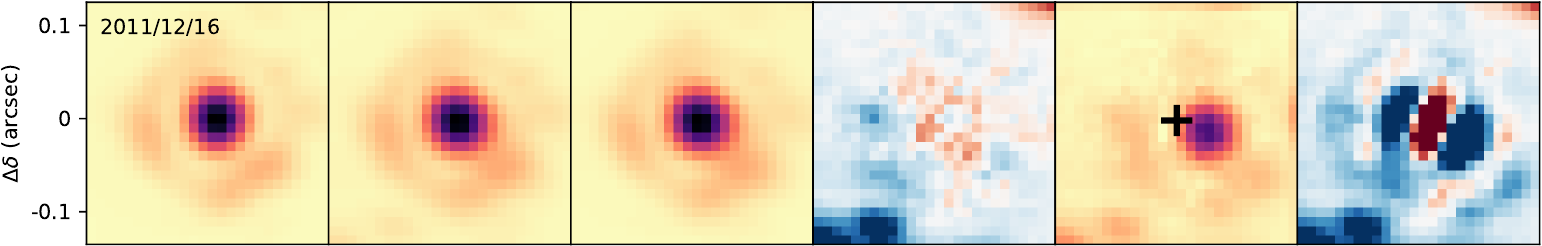}\\
   \includegraphics[width=0.75\textwidth]{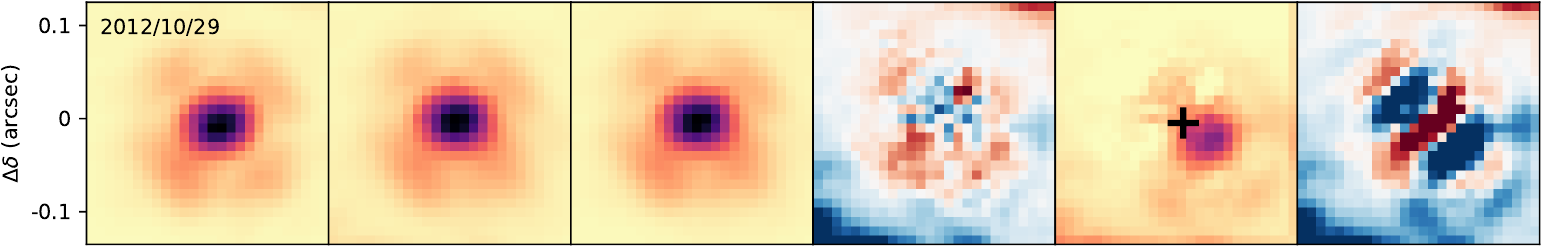}\\
   \includegraphics[width=0.75\textwidth]{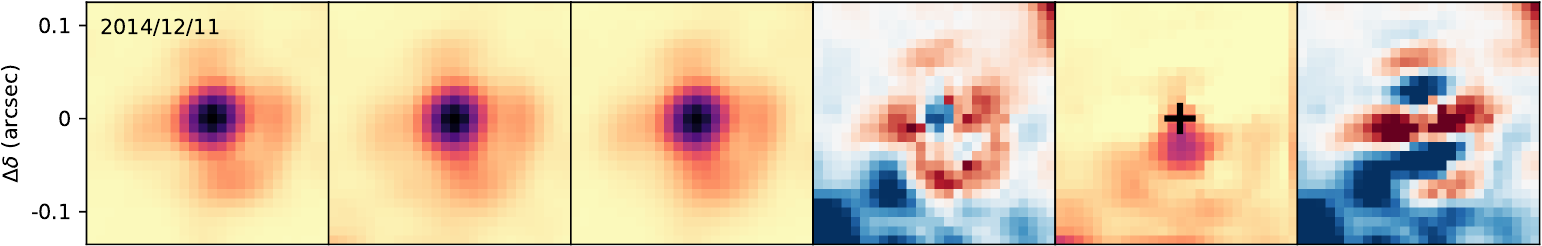}\\
   \includegraphics[width=0.75\textwidth]{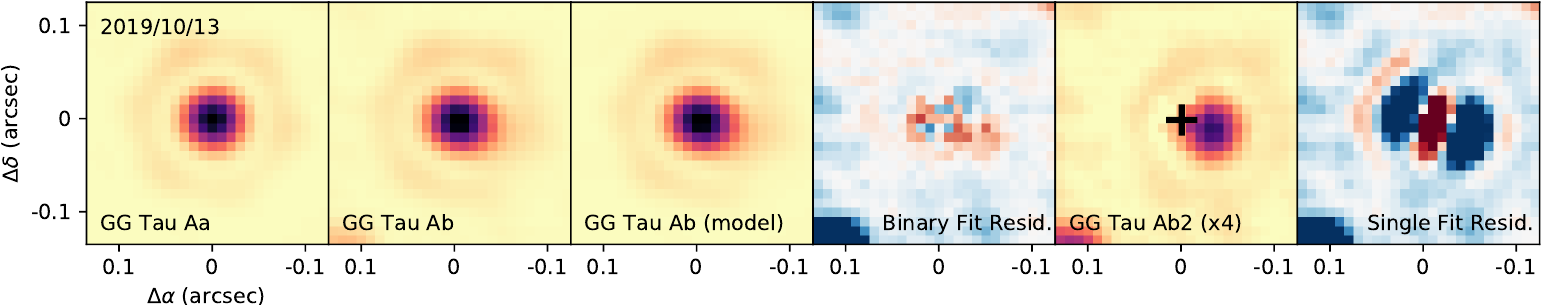}
      \caption{Thumbnails of Keck/NIRC2 images in the \ggt\ system. The first and second columns represent \ggt\,Aa and \ggt\,Ab, normalized to their peak values, respectively. The third column is the best-fit model of \ggt\,Ab created by shifting and scaling two copies of \ggt\,Aa, and the fourth column is the residuals from this model. Red and blue regions are negative and positive residuals, respectively, with the color map saturating at $\pm$3\% of the peak pixel in the image of \ggt\,Ab. The fifth column shows the result of subtracting the model of \ggt\,Ab1 (centered on the black cross) from the image of \ggt\,Ab, revealing \ggt\,Ab2, which is scaled up by a factor of four for visual purposes. The last column shows the residuals of fitting \ggt\,Ab as a single star instead; they are displayed with the same saturation levels as in the fourth column for direct comparison.}
         \label{fig:ggtau_detect}
   \end{figure*}

\ggt\ has been imaged at 2\,$\mu$m with the NIRC2 instrument behind the Keck\,II adaptive optics system at five epochs from 2010 to 2019 (PIs: A. Ghez, J.-L. Margot, J. Lu). In all cases, a number of short exposures were used to avoid saturation on the bright \ggt\,A. We retrieved the pipeline-calibrated data, subtracted a global frame median to remove any residual background, and performed image alignment and median combination to produce "epoch average" frames. Because the separation of the \ggt\,Aa--Ab pair is only $\approx$0\farcs25, much smaller than the atmospheric anisoplanetism angle, and assuming that it is a point source, we can consider \ggt\,Aa as a perfectly matched PSF that can be used to test whether the Ab component is itself resolved as a close pair. In 2010, we do not see any evidence that \ggt\,Ab is resolved, indicating that its separation and/or flux ratio was too small. In all other epochs, there is evidence for deviation from a single point source. The presence of the Ab2 component is most conspicuous in the 2019 data at a position angle of $\approx250$\degr\ but is also perceivable in the other epochs (see Figure\,\ref{fig:ggtau_detect}).

   \begin{figure*}
   \centering
   \includegraphics[width=0.46\textwidth]{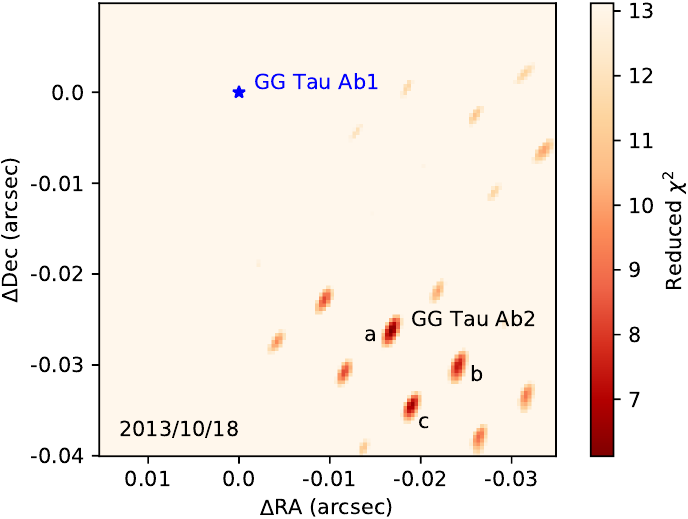}\hspace{0.3cm}
   \includegraphics[width=0.46\textwidth]{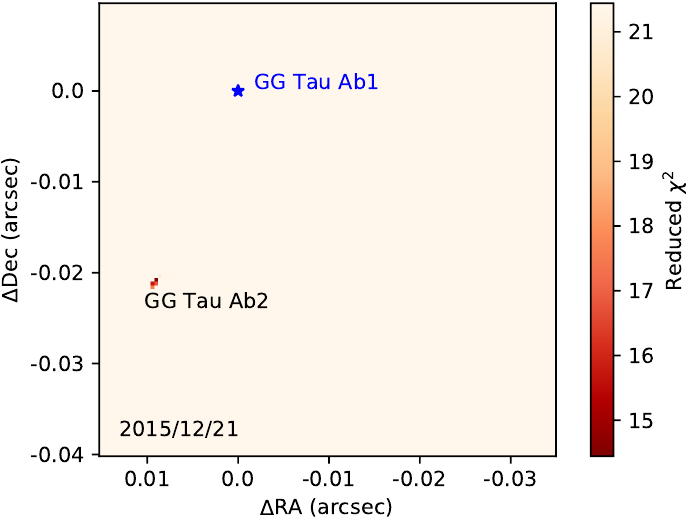}   
   \caption{Reduced $\chi^2$ maps for the location of \ggt\,Ab2 in the new PIONIER observations presented here. In the 2013 epoch, solution "a" is both the most likely from a pure likelihood perspective, and it is also the most plausible based on the motion observed between the 2012 and 2014 Keck observations (see Section\,\ref{subsec:obs_gg}). The second and third highest likelihood solutions (labelled "b" and "c"), located to the West-Southwest of the primary solution, are also illustrated in Figure\,\ref{fig:astrom}.}
         \label{fig:ggtau_pionier}
   \end{figure*}

To estimate the relative astrometry of the system, we performed a non-linear least square fit using the Levenberg-Marquart algorithm as implemented in {\tt lmfit} in which we modeled \ggt\,Ab as the linear combination of two translated and scaled copies of \ggt\,Aa. While the quality of the adaptive optics correction varies from epoch to epoch, and the fit is consequently sensitive to the initial guesses, we find that this approach is robust by performing the fit at the individual frame level (rather than on the epoch average frame) and use the standard deviation of the mean of these individual fits as an estimate of the random errors. Furthermore, we ran the binary fit with a range of initial parameters and used the range of resulting parameter values to evaluate the systematic uncertainties. In most cases, the latter dominate over the random uncertainties, and we consistently adopted the largest of the random and systematic uncertainties. This fitting process also yields the flux ratio of the system. In the case of a close asymmetric pair, however, the flux ratio and separation are generally correlated. This source of uncertainty is encapsulated in the methodology described above. As a consistency check, we also fitted the image of \ggt\,Ab as a single star; the resulting residuals are shown in the last column of Figure\,\ref{fig:ggtau_detect}. As expected from fitting a binary system as a single star, they contain a strong positive-negative-positive pattern along the binary position angle, further confirming the binary nature of the system.

Interferometric $H$-band observations with PIONIER were obtained in 2013 and 2015 (PI: E. Di Folco) to follow up on the 2012 discovery observations. Two datasets were obtained in 2015, but only the 2015/12/21 was of sufficient quality. The data were reduced using the PIONIER pipeline, similar to the results presented in \citet{difolco14}. We then fitted a model consisting of two point sources to the calibrated visibilities and phases. The resulting $\chi^2$ maps are shown in Figure\,\ref{fig:ggtau_pionier}. We note that the fits are imperfect in both epochs ($\chi^2_\mathrm{red} > 1$) due to contamination from the nearby \ggt\,Aa, but we consider that the relative astrometry of the close \ggt\,Ab pair is well determined nonetheless. In both the published 2012 and the new 2013 observations, PIONIER was set in its broadband mode, resulting in a very sparse sampling of the (u,v) plane and, consequently, in ambiguous results. Two possible solutions from the 2012 data were reported by \citet{difolco14} and we find three plausible solutions in 2013. In 2012, we could resolve the ambiguity by comparison with the nearly contemporaneous (2\,d difference) Keck observations. We find that the slightly less likely "second" solution matches very well with our Keck result whereas the most likely solution differs in position angle by 10\degr\ (see Figure\,\ref{fig:astrom}), a large enough deviation that we can confidently rule it out. In 2013, two of the three possible solutions have a separation of $\approx$0\farcs039 while the third has a separation of 0\farcs031. The latter is visually much more consistent with the orbital motion observed in the Keck images between 2012 and 2014\footnote{Both of the alternative solutions are also readily excluded by orbit fits that includes either of them instead of the nominal solution that we adopt here (see Section\,\ref{subsec:orbits}).}. We therefore adopted the latter solution. In 2015, PIONIER was used in its spectrally dispersed mode, producing a significantly improved (u,v) coverage and there is only one possible binary solution, which we adopt. We assigned the same astrometric uncertainties to both new PIONIER epochs, as systematic uncertainties are likely to dominate the error budget of such measurements. In principle, this analysis also yields an estimate of the binary flux ratio. Unfortunately, the presence of the nearby \ggt\,Aa components introduces a systematic offset in the measured correlated fluxes, which translates to a large systematic error on flux ratio. We could not properly assess this effect without independent simultaneous measurement of the \ggt\,Aa--Ab flux ratio and, therefore, could not reliably determine the flux ratio of \ggt\,Ab itself from the PIONIER data.

% - - - - - - - - - - - - - - - - - - - - - - - - - - - - - - - - - -
\subsection{\uxt\,B}
\label{subsec:uxt}

\begin{figure}
   \centering
   \includegraphics[width=\columnwidth]{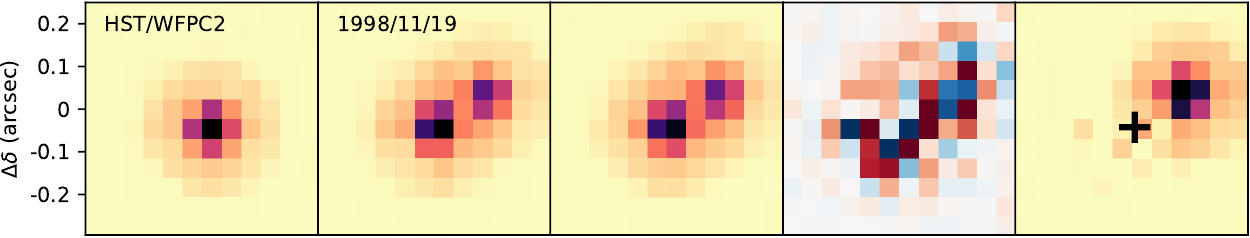}\\
   \includegraphics[width=\columnwidth]{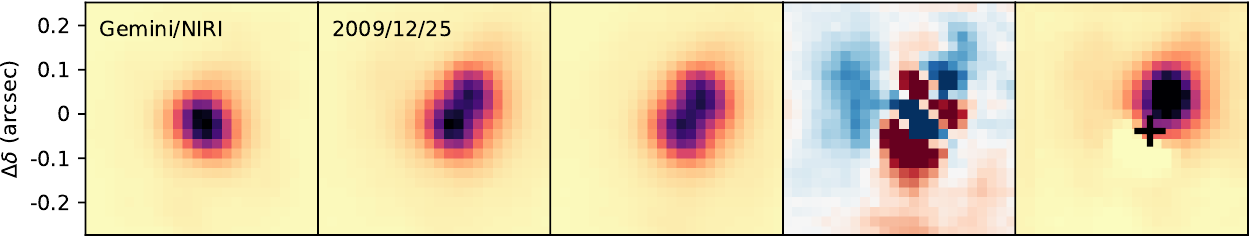}\\
   \includegraphics[width=\columnwidth]{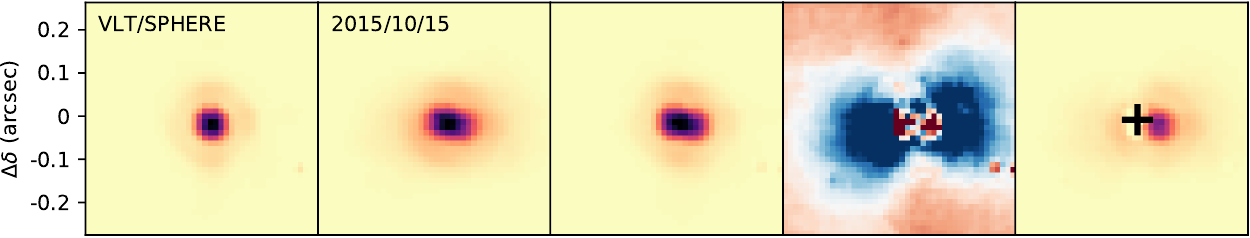}\\
   \includegraphics[width=\columnwidth]{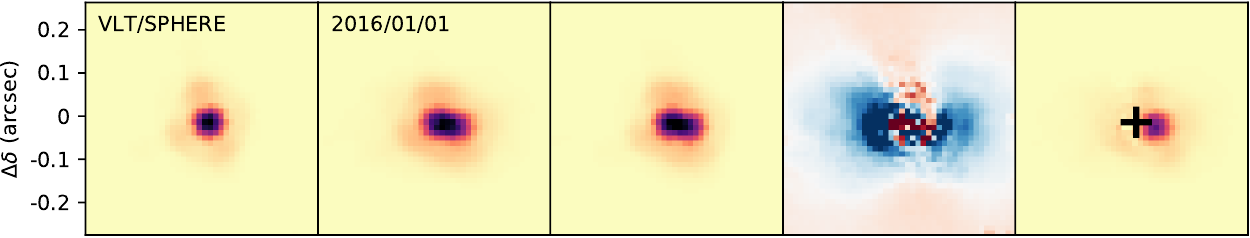}\\
   \includegraphics[width=\columnwidth]{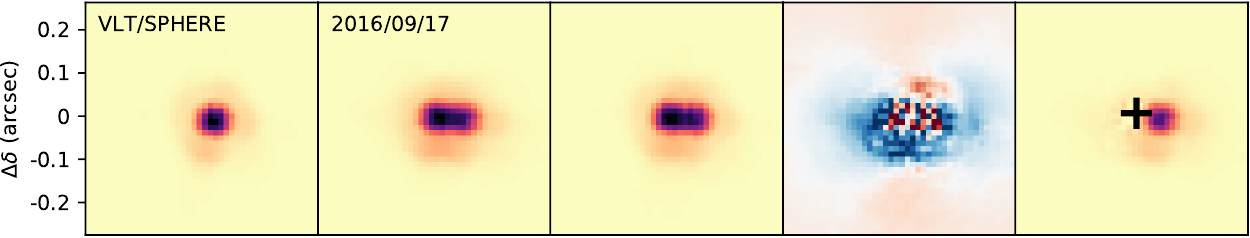}\\
   \includegraphics[width=\columnwidth]{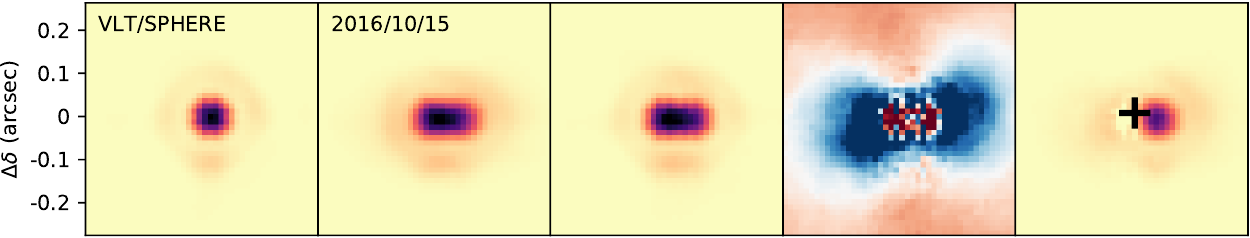}\\
   \includegraphics[width=\columnwidth]{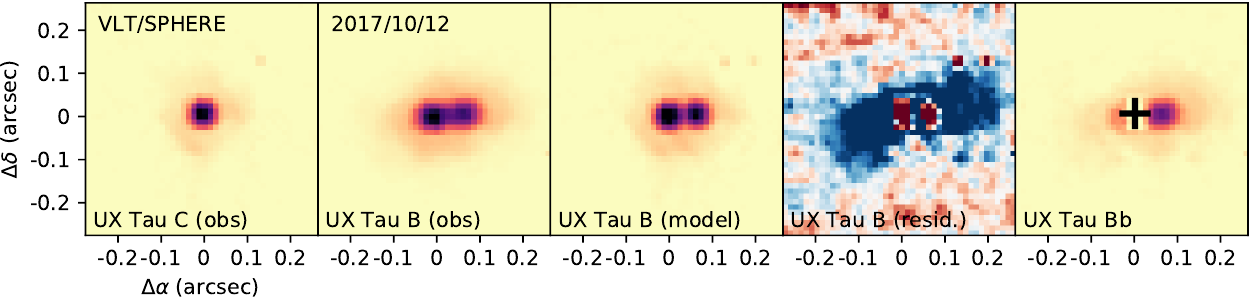}    
   \caption{Thumbnails of HST/WFCP2, Gemini/NIRI and VLT/SPHERE images in the \uxt\ system. The columns are similar to those in Figure\,\ref{fig:ggtau_detect}, except that no flux scaling is applied to \uxt\,Bb in the last column.}
         \label{fig:uxtau_detect}
   \end{figure}

The projected separation of the \uxt\,B pair was $\approx$0\farcs14 in its 1997 discovery epoch \citep{duchene99}, enabling multiple follow-up observations up to 2013 \citep{correia06, schaefer14}. We found two unpublished observations of the system in that same period in which the pair is clearly resolved, using HST/WFPC2 and Gemini/NIRI in 1997 and 2009, respectively (PIs: K. Stapelfeldt and P. Allen). We performed a similar analysis to the Keck/NIRC2 images of \ggt\ discussed above to estimate the relative astrometry of the \uxt\,B pair. Results are illustrated in the top two rows of Figure\,\ref{fig:uxtau_detect}). The HST/WFPC2 observations were obtained with two filters, which provide consistent estimates of the binary astrometry, but find that the flux ratio is wavelength-dependent: it rises from 0.5 to 0.64 between 0.6 and 0.8\,$\mu$m.

More recently, \uxt\ was observed at multiple epochs in 2015--17 with the VLT/SPHERE adaptive optic systems using the IRDIS dual-channel mode (PIs: J.-L. Beuzit, M. Benisty). We discarded one epoch in which the quality of the observations were of insufficient quality and again performed a similar binary fit as above. In most SPHERE observations \uxt\,A is either saturated or placed behind a coronagraphic mask and we therefore used \uxt\,C as estimate of the contemporaneous PSF. As shown in Figure\,\ref{fig:uxtau_detect}, the model residuals in some epochs are not negligible and elongated. We interpret this as evidence that the distance between the \uxt\,C and \uxt\,B is sufficient for deviations in the PSF to be present. Indeed, the position angle of the B--C pair is $\approx115\degr$, roughly aligned with the PSF elongation. Another possibility is that the \uxt\,C is marginally resolved due to scattering off its circumstellar dust, which was found to be elongated roughly along the North-South direction \citep{menard20}. Either way, as \uxt\,B is generally well-resolved, we trust that the binary astrometry is well determined with the systematic uncertainties that we estimated using the same method as for \ggt\,Ab. 

\begin{table*}
\caption{Astrometry of the \ggt\,Ab and \uxt\,B systems. $^\dagger$ Only the most plausible solutions from these epochs are listed here (see Section\,\ref{subsec:uxt}). The flux ratio of the \ggt\,Ab pair could not be determined in the new PIONIER data due to the contamination by the nearby \ggt\,Aa component.}
\label{tab:astrom}
\centering
\begin{tabular}{ccccccc}
\hline
Date & Instrument & Filter & Sep. (\arcsec) & PA (\degr) & Flux Ratio & Ref. \\
\hline
\multicolumn{7}{c}{\ggt\,Ab}\\
\hline
2011/12/16 & Keck/NIRC2 & Br$\gamma$ & 0.0305$\pm$0.0015 & 246.8$\pm$1.5 & 0.21$\pm$0.04 & This work \\
2012/10/29 & Keck/NIRC2 & $K'$ & 0.0347$\pm$0.0015 & 232.1$\pm$1.5 & 0.17$\pm$0.04 & This work \\
2012/10/31 & VLTI/PIONIER & $H$ & 0.0325$\pm$0.0004 & 230.0$\pm$0.4 & 0.23$\pm$0.10 & \citet{difolco14}$^\dagger$ \\
2013/10/18 & VLTI/PIONIER & $H$ & 0.0312$\pm$0.0004 & 213.2$\pm$0.4 & (\dots) & This work$^\dagger$ \\
2014/12/11 & Keck/NIRC2 & $K'$ & 0.0274$\pm$0.0020 & 181.5$\pm$2.0 & 0.15$\pm$0.03 & This work \\
2015/12/21 & VLTI/PIONIER & $H$ & 0.0227$\pm$0.0003 & 157.3$\pm$0.4 & (\dots) & This work \\
2019/10/13 & Keck/NIRC2 & Br$\gamma$  & 0.0315$\pm$0.0010 & 252.3$\pm$1.0 & 0.22$\pm$0.03 & This work \\
\hline
\multicolumn{7}{c}{\uxt\,B}\\
\hline
1997/12/12 & CFHT/AOBIR & $K$ & 0.138$\pm$0.002 & 303.9$\pm$1.0 & 0.78$\pm$0.05 & \citet{duchene99} \\
1998/11/19 & HST/WFPC2 & F606W & 0.142$\pm$0.002 & 302.1$\pm$1.0 & 0.50$\pm$0.02 & This work \\
2002/10/22 & VLT/NACO & $K$ & 0.136$\pm$0.002 & 309.0$\pm$1.0 & 0.81$\pm$0.02 & \citet{correia06} \\
2009/10/25 & Keck/NIRC2 & $K_c$ & 0.084$\pm$0.002 & 326.9$\pm$1.0 & 0.81$\pm$0.02 & \citet{schaefer14} \\
2009/12/25 & Gemini/NIRI & $K_c$ & 0.085$\pm$0.002 & 327.4$\pm$1.0 & 0.78$\pm$0.01 & This work \\
2011/10/12 & Keck/NIRC2 & $K_c$ & 0.056$\pm$0.002 & 339.7$\pm$1.0 & 0.80$\pm$0.01 & \citet{schaefer14} \\
2013/01/27 & Keck/NIRC2 & $K_c$ & 0.032$\pm$0.005 & 359.5$\pm$8.7 & 0.9$\pm$0.4 & \citet{schaefer14} \\
2015/10/25 & VLT/SPHERE & $H$ & 0.047$\pm$0.002 & 259.4$\pm$1.0 & 0.82$\pm$0.03 & This work \\
2016/01/01 & VLT/SPHERE & $H$ & 0.046$\pm$0.002 & 259.1$\pm$1.0 & 0.83$\pm$0.04 & This work \\
2016/09/17 & VLT/SPHERE & $H$ & 0.057$\pm$0.002 & 264.9$\pm$1.0 & 0.82$\pm$0.02 & This work \\
2016/10/15 & VLT/SPHERE & $K_c$ & 0.061$\pm$0.003 & 269.0$\pm$1.0 & 0.83$\pm$0.01 & This work \\
2017/10/12 & VLT/SPHERE & $H$ & 0.070$\pm$0.002 & 272.8$\pm$1.0 & 0.79$\pm$0.03 & This work \\
\hline
\end{tabular}
\end{table*}

%--------------------------------------------------------------------
\section{Results}

% - - - - - - - - - - - - - - - - - - - - - - - - - - - - - - - - - -
\subsection{Relative astrometry and orbital fits}
\label{subsec:orbits}

\begin{figure*}
   \centering
   \includegraphics[width=0.49\textwidth]{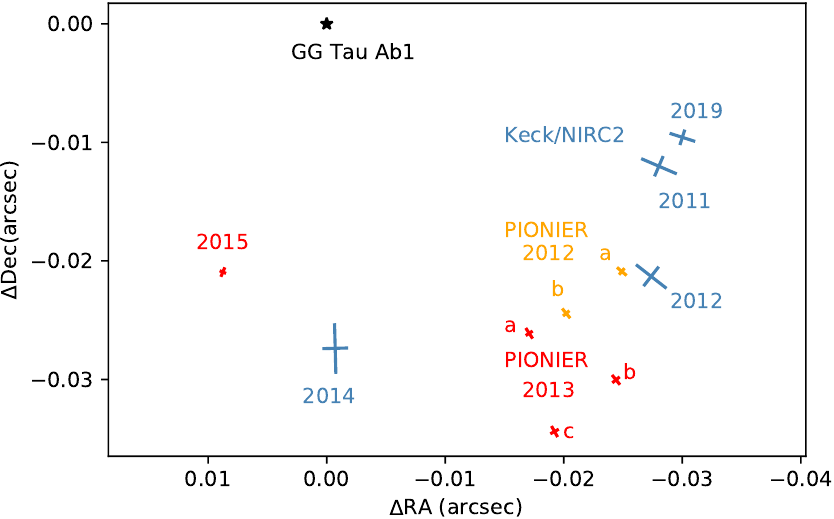}
   \includegraphics[width=0.49\textwidth]{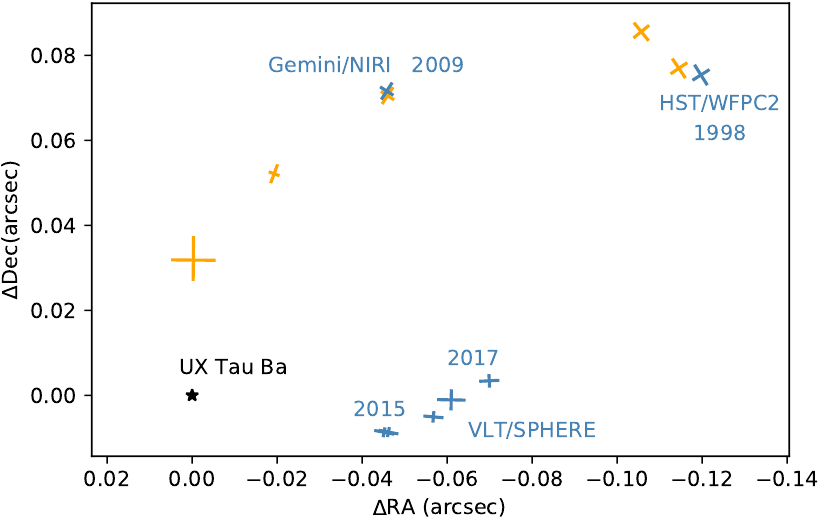}
   \caption{Relative orbit of the \ggt\,Ab and \uxt\,B binaries. In both panels, orange symbols represent published astrometric datapoints whereas blue symbols indicate new adaptive optics or HST observations that are presented here for the first time. In 2012 and 2013 (orange and red symbols, respectively), the PIONIER observations of \ggt\,Ab allowed two and three possible solutions, respectively. Based on the Keck observations, solutions "a" in both epochs are the most plausible ones. }
   \label{fig:astrom}
   \end{figure*}

All published and new astrometric results are represented in Table\,\ref{tab:astrom} and illustrated in Figure\,\ref{fig:astrom}. In both cases, the new observations presented here represent a significant improvement in the orbital coverage of both systems. For \ggt\,Ab, not only do we present the first observation of orbital motion, but the system has nearly completed a full orbit between 2011 and 2019. Indeed, this $\approx$8\,yr period further confirms the tentative "a posteriori detection" in 2003 discussed by \citet{difolco14} if we adopt a position angle of $\approx$245\degr\ for that epoch. In the case of \uxt\,B, while the position angle of the system had changed by $\approx$60\degr\ between 1998 and 2013, the new SPHERE observations represent a change of $\approx$270\degr\ in the subsequent 2\,yr, and rapid year-by-year motion is evident between 2015 and 2017. These most recent observations lie at intermediate position between the 25 and 40\,yr-period orbits presented by \citet{schaefer14}. This improved coverage invites orbital fits, which we expect to be well constrained by existing observations.

\begin{figure*}
   \centering
   \includegraphics[width=0.49\textwidth]{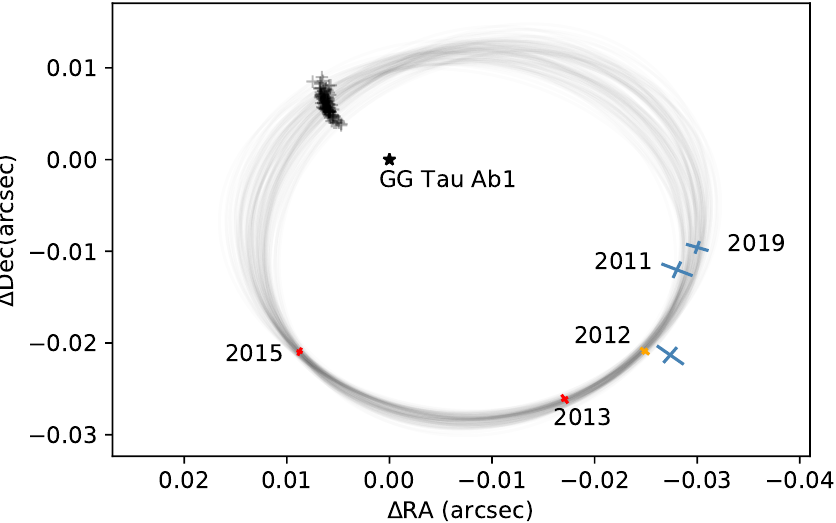}
   \includegraphics[width=0.49\textwidth]{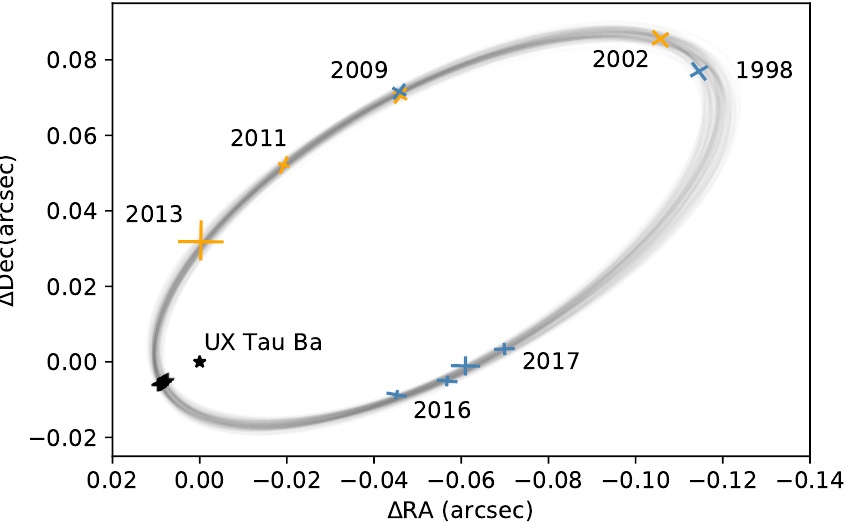}
   \caption{Random set of 100 orbital solutions selected form the converged part of the MCMC chain for the \ggt\,Ab and \uxt\,B binaries (considering the "trimmed" dataset, see Section\,\ref{subsec:orbits}). Black $+$ symbols indicate the periastron of each plotted orbit.}
   \label{fig:orbits}
   \end{figure*}

To perform the orbital fit, we used the {\tt orbitize!} package \citep{blunt19} in combination with the parallel-tempered package {\tt ptemcee} \citep{vousden16}. Given the tentative nature of the 2003 detection from \citet{difolco14}, and our lack of success in detecting \ggt\,Ab2 in VLT/NaCo images at any epoch, we decided against using it in the fit. We nonetheless verified that, given its poor astrometric precision, including it would not significantly alter the fit. The free parameters used in {\tt orbitize!} are the eccentricity ($e$), inclination ($i$), phase of periastron ($\tau_0$, measured relative to the earliest data point), position angle of the line of nodes ($\Omega$), argument of periastron ($\omega$), semi-major axis ($a$), and total system mass ($M_{sys}$). During the fit, we fixed the parallax of the system to $\pi = 6.91$\,mas \citep[the median L1551 estimate from][see Section\,\ref{subsec:mdyn}]{galli19} to convert from angular to physical separations, and we further converted $M_{sys}$ into the orbital period ($P$) and $\tau_0$ into the time of passage at periastron, $T_0$. These three quantities are those readily constrained by the data. Uniform priors are used for $e$, $\omega$, $\Omega$ and $\tau_0$, log-uniform and sine-uniform are used for $a$ and $i$, respectively, and a Gaussian prior is used for $M_{sys}$. In the absence of radial velocity measurements, solutions defined by ($\omega$, $\Omega$) and ($\omega+180\degr$, $\Omega+180\degr$) are degenerate; when extracting the MCMC chains, we solve the ambiguity by enforcing that $0\degr \leq \Omega < 180\degr$. We applied broad priors on $M_{sys}$ that were informed by preliminary estimates of the semi-major axis and orbital period. Specifically, we used 1.0$\pm$0.5\,$M_\odot$ and 2$\pm$1\,$M_\odot$ for \ggt\,Ab and \uxt\,B, respectively. We performed each fit with ten temperatures and 100 walkers per temperature, and advanced the chains 40000 steps past an initial 1000 steps of burn-in, and only kept every tenth walker position to remove correlations in the chains. Inspection of the chain evolution confirms that they have reached convergence. The resulting cornerplots are presented in Appendix\,\ref{sec:appendix} and 68-percentile for all parameters are listed in Table\,\ref{tab:orbits}. We note that the large orbital coverage is such that there is no strong correlation between orbital parameters, with the exception of $\Omega$ and $\omega$. The location of the periastron is nonetheless well constrained in both cases (see Figure\,\ref{fig:orbits}). Finally, we note that we have adopted the astrometric solution with $i\leq90\degr$ instead of the solution with $i>90\degr$, as justified in Section\,\ref{subsec:architecture}, even though the data at hand do not allow to solve that ambiguity.

The orbital fits for both systems are reasonable but, as indicated in Table\,\ref{fig:orbits}, neither of the best orbital fit reaches $\mathrm{min}(\chi^2_\mathrm{red}) = 1$, suggesting some underestimates of the astrometric uncertainties and/or the presence of outliers. We inspected the fit residuals and determined that the 2014 measurement of \ggt\,Ab and the 1997, 1998 and 2015 measurements of \uxt\,B were the most discrepant (at the $\approx2.5-3\sigma$ level). We therefore repeated the orbital fit with "trimmed" datasets, namely excluding the 2014 measurement of \ggt\,Ab, and the 1998\footnote{We consider the 1997 HST datapoint more robust than the discovery 1998 CFHT measurement from \citet{duchene99} as precise astrometry was not a priority in that study.} and 2015 measurements of \uxt\,B. In both cases, the fits to the "trimmed" dataset produces a much better fit, with $\mathrm{min}(\chi^2_\mathrm{red}) \approx 1.0-1.3$. The confidence intervals for the "complete" and "trimmed" dataset overlap at the $1\sigma$ level, confirming that no datapoint biases the fit significantly. In most cases, the resulting uncertainties on orbital parameters are comparable for both fits, with the main exception being the semi-major axis and, consequently, system mass for \ggt\,Ab. We conservatively adopt the "trimmed" dataset fit for the subsequent analysis.

\begin{table*}
\caption{Orbital parameters of the \ggt\,Ab and \uxt\,B systems. Parameters listed in the lower part of the table (periastron and apastron distances, and total system masses) are derived from the seven free parameters of the orbital fit in combination with the estimate of the distance to the system. Consequently, two uncertainties are listed, respectively the random (resulting from the orbit fit) and systematic (driven by the distance estimate) uncertainties. ``Trimmed" datasets are those in which the most discrepant astrometric points from the fit to ``Complete" dataset are discarded prior to the orbital fit (see Section\,\ref{subsec:orbits}).}
\label{tab:orbits}
\centering
\begin{tabular}{c|cc|cc}
\hline
 & \multicolumn{2}{c|}{\ggt\,Ab} & \multicolumn{2}{c}{\uxt\,B} \\
 & "Complete" dataset & "Trimmed" dataset & "Complete" dataset & "Trimmed" dataset \\
\hline
  $P$ (yr) & 8.16\,$^{+0.08}_{-0.07}$ & 8.13\,$^{+0.08}_{-0.07}$ & 32.0\,$^{+0.9}_{-0.8}$ & 31.9\,$^{+1.0}_{-0.9}$ \\
  $T_0$ & 2017.59\,$^{+0.28}_{-0.25}$ & 2017.14\,$^{+0.24}_{-0.21}$ & 2013.89\,$^{+0.14}_{-0.13}$ & 2013.95\,$^{+0.16}_{-0.14}$ \\
  $e$ & 0.539\,$^{+0.033}_{-0.027}$ & 0.574\,$^{+0.048}_{-0.040}$ & 0.870\,$^{+0.010}_{-0.009}$ & 0.863\,$\pm$\,0.011 \\
  $a$ (mas) & 25.7\,$^{+1.0}_{-0.8}$ & 26.2\,$^{+2.1}_{-1.4}$ & 79.1\,$^{+2.4}_{-2.2}$ & 78.8\,$^{+2.6}_{-2.1}$ \\
%  $i$ (\degr) & 139.9\,$^{+4.5}_{-4.4}$ & 140.1\,$^{+5.6}_{-6.0}$ & 32.2\,$^{+5.0}_{-7.2}$ & 33.5\,$^{+4.9}_{-6.8}$ \\
  $i$ (\degr) & 40.1\,$^{+4.4}_{-4.5}$ & 39.9\,$^{+6.0}_{-5.6}$ & 32.2\,$^{+5.0}_{-7.2}$ & 33.5\,$^{+4.9}_{-6.8}$ \\
  $\Omega$ (\degr) & 100.8\,$^{+14.4}_{-11.2}$ & 125.9\,$^{+9.4}_{-15.3}$ & 155.1\,$^{+8.1}_{-9.2}$ & 152.0\,$^{+8.7}_{-10.4}$ \\
%  $\omega$ (\degr) & 67.3\,$^{+7.8}_{-5.0}$ & 82.3\,$^{+5.1}_{-9.3}$ & 321.8\,$^{+9.8}_{-9.2}$ & 325.2\,$^{+11.3}_{-10.1}$ \\
  $\omega$ (\degr) & 292.7\,$^{+5.0}_{-7.8}$ & 277.7\,$^{+9.3}_{-5.1}$ & 321.8\,$^{+9.8}_{-9.2}$ & 325.2\,$^{+11.3}_{-10.1}$ \\
  \hline
  $d_\mathrm{peri}$ (au) & 1.70\,$^{+0.09}_{-0.08}\,\pm\,0.04$ & 1.62\,$^{+0.06}_{-0.08}\,\pm\,0.04$ & 1.49\,$\pm\,0.11\,\pm$\,0.03 & 1.56$\,\pm\,$0.12$\,\pm\,$0.04 \\
  $d_\mathrm{apo}$ (au) & 5.73\,$^{+0.31}_{-0.26}\,\pm\,0.13$  & 5.97\,$^{+0.68}_{-0.45}\,\pm\,0.14$ & 21.4\,$^{+0.7}_{-0.6}$$\,\pm$\,0.5 & 21.2\,$^{+0.8}_{-0.6}$$\,\pm$\,0.5 \\
  $M_{sys} (M_\odot)$ & 0.77\,$^{+0.09}_{-0.07}\,\pm\,0.05$ & 0.82\,$^{+0.22}_{-0.12}\,\pm\,0.06$ & 1.47\,$^{+0.15}_{-0.14}\,\pm\,$0.10 & 1.47\,$^{+0.15}_{-0.14}\,\pm\,$0.10 \\
\hline
  min($\chi^2_\mathrm{red}$) & 2.40 & 1.26 & 1.75 & 1.03 \\
\hline
\end{tabular}
\end{table*}

% - - - - - - - - - - - - - - - - - - - - - - - - - - - - - - - - - -
\subsection{Dynamical masses}
\label{subsec:mdyn}

\begin{figure}
   \centering
   \includegraphics[width=\columnwidth]{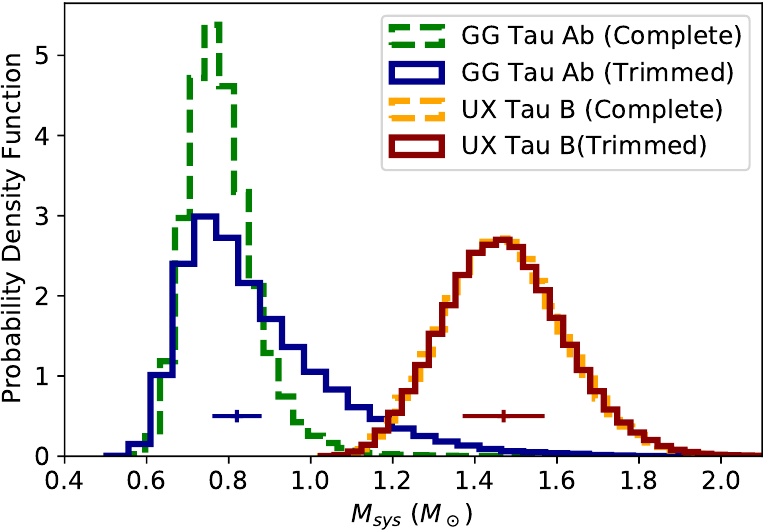}
   \caption{Posterior distribution for the dynamical mass of the \ggt\,Ab and \uxt\,B systems. In both cases, dashed and solid histograms represent the distributions using "complete" and "trimmed" astrometric datasets, respectively. The black horizontal bars represent the systematic error introduced by the uncertainty on the distance to each system.}
   \label{fig:mdyn}
\end{figure}

We then proceeded to convert the orbital solutions in terms of angular scale to physical scale, which requires determining the distance to the two systems. Specifically, $M_{sys} = a^3\, P^{-2}\, D^3\,M_\odot$, where $a$, $P$ and $D$ are expressed in arcsec, yr and pc, respectively. \ggt\,A is included in the GAIA DR3 catalog \citep{gaia16, gaia23} but with a very poor RUWE parameter, most likely because of the internal orbital motion of the Aa--Ab pair. \ggt\,Ba and \ggt\,Bb are in the GAIA DR3 catalog with good RUWE parameters, however. Their parallaxes are 6.87$\pm$0.04 and 6.75$\pm$0.15\,mas, respectively. Similarly, \uxt\,B has a bad RUWE parameter and its GAIA DR3 parallax cannot be trusted but \uxt\,A and C have good measurements, with parallaxes of 7.03$\pm$0.03 and 6.92$\pm$0.07\,mas, respectively. In both cases the available measurements are marginally consistent with each other. We also note that both \ggt\, and \uxt\ are part of the L1551 group, for which the mean parallax and associated dispersion is 6.91$\pm$0.16\,mas \citep{galli19}. Since the latter is consistent with all individual estimates and is more conservative, we adopted the L1551 parallax to determine the dynamical mass of both systems and consider that its associated uncertainty represents a systematic uncertainty, which we carry separately in Table\,\ref{tab:orbits} and illustrate in Figure\,\ref{fig:mdyn}. For both systems, the systematic uncertainty is commensurate with, but smaller than, the random uncertainty. 

% - - - - - - - - - - - - - - - - - - - - - - - - - - - - - - - - - -
\subsection{System architectures}
\label{subsec:architecture}

\begin{figure}
   \centering
   \includegraphics[width=\columnwidth]{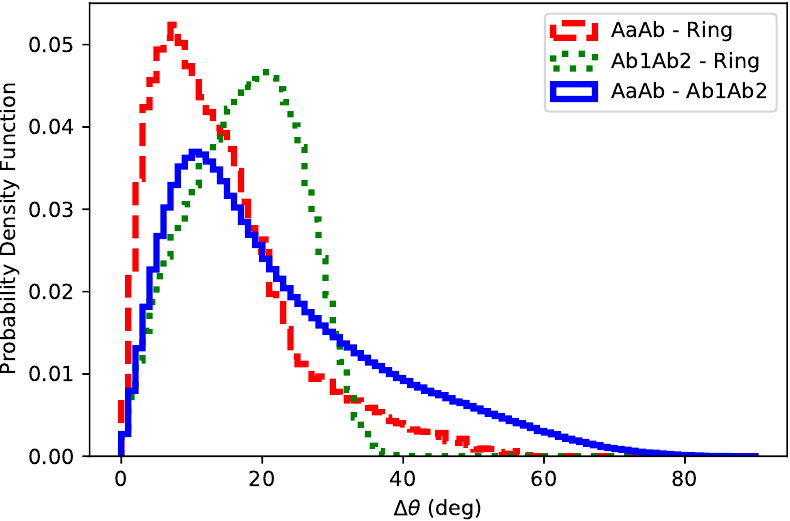}
   \caption{Angle between the orbital and ring angular momentum vectors in the \ggt system. In all cases, we solve the orbital inclination ambiguity by assuming that the planes are close to coplanar.}
   \label{fig:relorient}
\end{figure}

Since the orbit properties of both systems are well constrained, we can compare the relative orientation of their orbital plane with other relevant planes in each system (see Table\,\ref{tab:archi} and Figure\,\ref{fig:relorient}), as this is relevant to the stability and dynamical evolution of multiple systems and their associated disks. We computed the angle between two (orbital or disk) angular momentum vectors, $\Delta\theta$, following Equation\,1 in \citet{czekala19}, using Monte Carlo method to sample the input posterior distributions. We note that for visual orbits, there is an inclination ambiguity whereby orbits with $i$ and $180\degr-i$ inclinations (and, conjointly, $\omega$ and $360\degr-\omega$ argument of periastron) are indistinguishable in the plane of the sky. When comparing with other planes, we must therefore consider two alternative deprojection solutions. In the case of \ggt, we note that with inclinations of $\approx35-40\degr$, adopting the alternative inclination leads to orbits that are nearly perpendicular to each other. Similarly, the wider \ggt\,Aa--Ab orbit is either close to coplanar or nearly perpendicular to the outer ring. While circumbinary disk in a polar configuration and highly misaligned hierarchical triple systems exist \citep[e.g.,][]{tokovinin17, czekala19}, we focus here on the configurations that are close to coplanar. We also note that the relative orientation of the tighter orbit with respect to the circumtriple disk does not significantly affect the evolution of the system \citep{ceppi23}.

\begin{table*}
\caption{Angles defining key planes in the \ggt\ and \uxt\ systems.}
\label{tab:archi}
\centering
\begin{tabular}{ccccc}
\hline
System & Plane & $i\,(\degr)$ & $\Omega\,(\degr)$ & Reference \\
\hline
\ggt & Ab1--Ab2 orbit & 39.9\,$^{+6.0}_{-5.6}$ & 125.9\,$^{+9.4}_{-15.3}$ & This work \\
   & Aa--Ab orbit & 36.2\,$^{+6.4}_{-5.3}$ & 99.7\,$^{+19.2}_{-39.9}$& \cite{toci24} \\
   & Circumtriple ring & 35$\pm$2 & 97$\pm$1 & \cite{phuong20} \\
\hline
\uxt & B orbit & 33.5\,$^{+4.9}_{-6.8}$ & 152.0\,$^{+8.7}_{-10.4}$ & This work \\
   & A disk & 32$\pm$8 & 164$\pm$8 & \cite{zapata20} \\
\hline
\end{tabular}
\end{table*}

In the \ggt\,A system, we can compare the Ab1--Ab2 orbital solution we have obtained to the updated Aa--Ab orbit presented in \cite{toci24}. Because the latter is only modestly constrained, the resulting misalignment angle is only loosely constrained. The most plausible value is $\sim12\degr$, suggesting only a modest degree of misalignment between the two orbits. It is worth noting that the Aa--Ab pair has a distribution of periastron distances ($a(1-e)$) of $31^{+4}_{-5}$\,au, which is about five times larger than the apoastron distance of the Ab1--Ab2 pair. This ratio is relevant in the context of the dynamical stability of the system (see Section\,\ref{subsec:stability}). Another well-characterized plane in the \ggt\,A system is that of the massive dust ring that surrounds it. Excluding the near polar configurations, we find that both orbits have low relative inclinations to the ring, with peaks in the posteriors of $\Delta\theta\approx7\degr$ and 20\degr\ for the Aa--Ab and Ab1--Ab2 orbits, respectively.

Prior to this study, no orbit had been estimated in the \uxt\ system due to the long orbital period associated with the large projected separations, and correspondingly long orbital periods, of the A--C and A--B pairs. It has not even been demonstrated whether they are physically bound, although it is statistically likely \citep[e.g.,][]{leinert93}. The apoastron distance of the Ba--Bb pair we find, $\approx20$\,au, is $\approx40$ times smaller than the A--B projected separations. Regarding the orbital plane, the only other plane that is well determined in the system is that of the \uxt\,A disk \citep{francis20, zapata20, menard20}. Considering the two ambiguous solutions, we find that the \uxt\,Ba--Bb orbit is either mildly misaligned ($\Delta\theta \lesssim25\degr$ with a peak in the posterior at $\Delta\theta\approx10\degr$) with, or highly tilted ($\Delta\theta = 115\pm 25\degr$) relative to the \uxt\,A disk. 

%--------------------------------------------------------------------
\section{Discussion}
\label{sec:discuss}

% - - - - - - - - - - - - - - - - - - - - - - - - - - - - - - - - - -
\subsection{Stellar properties}

The dynamical masses derived in this study must be interpreted in light of past studies of both systems. Here we consider how our results modify our understanding of the stellar properties in these systems.

\ggt\ is among the best-studied T\,Tauri multiple systems, in large part because of the massive dust ring and gaseous disk that surrounds \ggt\,A. Its gas dynamics has allowed for precise estimates of the total dynamical mass of \ggt\,A \citep[e.g.,][]{guilloteau99}. Most recently, \citet{phuong20} derived a total system mass of 1.41$\pm$0.08\,$M_\odot$ (scaled to a distance of 145\,pc adopted here). Combined with our \ggt\,Ab dynamical mass, this implies that the mass of \ggt\,Aa is 0.59$^{+0.14}_{-0.22}\,M_\odot$, with the \ggt\ Ab subsystem being at least as massive as, and up to $\sim60\%$ more massive than, \ggt\,Aa. The spectral types for \ggt\,Aa and Ab have been estimated at K7--M0 and M0.5--M2, respectively \citep{white99, hartigan03}. Based on the observed range of dynamical masses for single stars in Taurus \citep{simon19, braun21, flores22}, we estimate that \ggt\,Aa is most consistent with a $\approx0.6\,M_\odot$ mass. Similarly, we propose that \ggt\,Ab1 and Ab2 have masses of $\approx0.5-0.6\,M_\odot$ and $\approx0.2-0.3\,M_\odot$, respectively. The observed $K$ band flux ratio of the Ab1--Ab2 pair, if interpreted as representative of the stellar photospheres, would be most consistent a 1:3 mass ratio based on the {\tt BT-Setll} models \citep{allard12}, but we caution that the presence of continuum emission from circumstellar material precludes reaching a definitive conclusion without spatially resolved spectroscopy.

In the absence of a circumstellar disk whose Keplerian motion could be used \citep[e.g.,][]{akeson19}, the dynamical mass estimate we obtained for \uxt\,B is the first for this component. Spectroscopic observations have led to a spectral type estimate of M1--M2 \citep{hartigan94, white01, luhman18}. Based on the compilation of Taurus dynamical masses mentioned above, this would suggest a 0.5--0.7\,$M_\odot$ mass for the individual masses, somewhat in tension with the dynamical mass of $\approx$1.5\,$M_\odot$ that we obtain. The systematic uncertainty introduced by the distance to the system does not appear large enough to account for this mismatch. We note, however, that the mass ratio of the close pair appears close to unity based on flux ratio measurements \citep{duchene99, correia06, schaefer14} and it is possible that a slightly earlier spectral type ($\approx$M0.5) for both stars would be consistent with the derived dynamical mass. 

% - - - - - - - - - - - - - - - - - - - - - - - - - - - - - - - - - -
\subsection{Orbital properties and circumstellar matter}

\begin{figure*}
   \centering
   \includegraphics[width=0.49\textwidth]{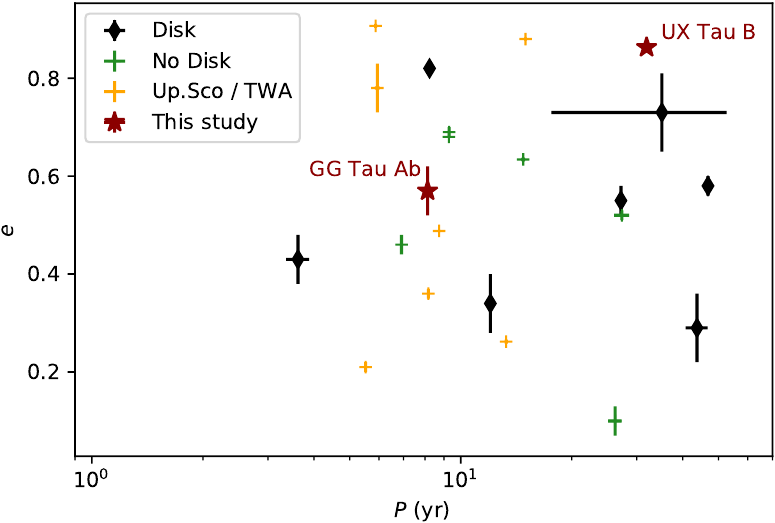}
   \includegraphics[width=0.49\textwidth]{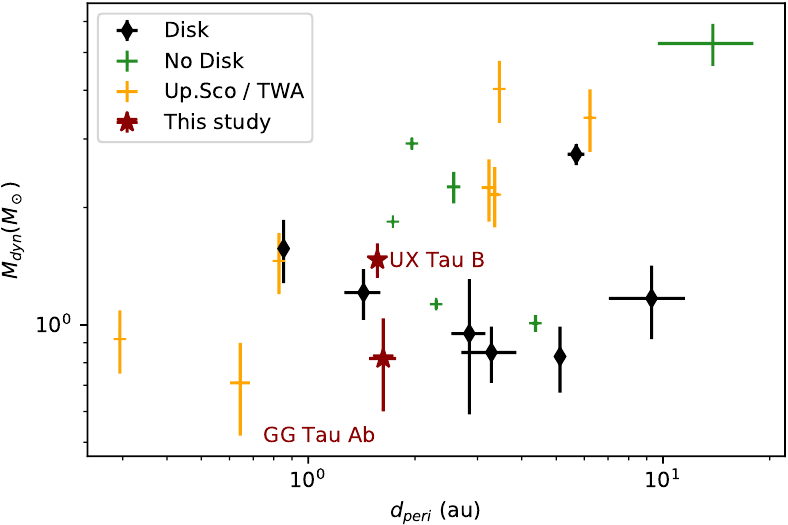}
   \caption{Orbital properties of T\,Tauri visual binaries. In both panels, black diamonds and green crosses represent $\lesssim3$\,Myr-old systems with and without circumstellar material, respectively, while yellow crosses are $\sim10$--20\,Myr-old disk-free systems in Upper Scorpius and the TW\,Hya association. The two red stars mark the systems studied here. {\it Left:} Comparison of orbital period and eccentricity based on literature results listed in Appendix\,\ref{sec:litt_orbits}. {\it Right:} Comparison of periastron distances and dynamical masses for the same systems.}
   \label{fig:p_e}
\end{figure*}

We now consider the implications of the orbital properties we have inferred for the two systems. We first note that, while \ggt\,Ab has a moderate eccentricity, \uxt\,B is among the highest eccentricity systems known among pre-main sequence visual binaries (see Figure\,\ref{fig:p_e}). Among main sequence binary systems, such a high eccentricity is also close to the maximum observed for this range of orbital periods \citep{raghavan10}. This limit is set by the onset of (partial) tidal circularization over the lifetime of the system. Therefore, the system could survive if its current configuration as far as interactions between individual stars are concerned.

The complete orbital solutions we have found place stringent constraints on where gas and dust can be maintained in a stable configuration. In particular, the similar $\approx$1.6\,au periastron distances for both systems imply that circumstellar disks around each star can be no larger than $\lesssim1$\,au. No infrared excess is detected in \uxt\,B but there is one in \ggt\,Ab, which is also found to be accreting \citep{white01,mccabe06}. Irrespective of which of the two components hosts this material, this proves that it is possible for extremely compact disks to survive in close binaries, despite the general tendency against it \citep[][see also Figure\,\ref{fig:p_e}]{cieza09, kraus12} or the expectation of a very short viscous timescale \citep[$\lesssim0.05$\,Myr,][]{cieza09}. Indeed, there is at least one system with an even smaller periastron distance, similar eccentricity, and confirmed infrared excess \citep[ROXs\,47,][]{rizzuto16}. The presence of streamers from the massive outer ring onto both \ggt\,Aa and Ab \citep{keppler20} confirms that continuous feeding of (small) circumstellar disks is likely ongoing, providing a possible explanation to the disk survival and on-going accretion. We speculate that the lack of circumstellar material around either component of \uxt\,B is due to a lack of outer reservoir that can feed the system, possibly compounded by its high eccentricity. Indeed, the higher binary eccentricity leads to regions of stable orbits around  either component that are simply too small to sustain long-lived disks.

We find that the two orbital planes in the \ggt\ system have a modest degree of misalignment ($\Delta\theta\sim5-20\degr$) with each other and with the circumtriple ring, in line with a scenario in which the entire system formed from a single core undergoing collapse and fragmentation \citep{offner23}. It is worth pointing out that the naive expectation would be for the circumtriple ring to be coplanar with the invariable plane of the triple system, which is distinct from the two orbital plane themselves, especially so since the Ab subsystem is somewhat more massive than the Aa component. We also stress that without radial velocity measurements or additional dynamical considerations (size of the circumtriple ring, three-body stability), we cannot exclude the orbital solutions in which the ring is on a roughly polar orbit around the Aa--Ab system.

% - - - - - - - - - - - - - - - - - - - - - - - - - - - - - - - - - -
\subsection{System stability and dynamics}
\label{subsec:stability}

Hierarchical triple systems are prone to disruptive instability if their orbital properties are such that all three star-star separations become commensurable at some point along the orbits. In order to remain stable, it is therefore necessary that the tightest pair be much more compact than the separation to the outer third component. This has been extensively empirically verified for main sequence systems \citep[e.g.,][]{tokovinin14, tokovinin21}. Furthermore, numerical studies have explored this boundary, taking into account orbital eccentricity and deviation from coplanarity and over a broad range of mass ratios \citep{eggleton95, mardling01, he18}. With its large projected separation relative to the semi-major axis of the close binary, the \uxt\,A--B system is comfortably in a stable configuration. The case of \ggt\,A is not as extreme, yet both its ratio of orbital periods ($\gtrsim25$) and the ratio of periastron of the outer orbit to periastron of the inner orbit (see Section\,\ref{subsec:architecture}) place it in the stable regime \citep{he18} and among ``typical" systems \citep{tokovinin21}. We note, however, that this assessment assumes that the evolution of the system is driven exclusively by $N$-body dynamics, whereas continued accretion of gas and the interaction with the massive outer ring could further modify either orbits. Therefore, the long-term stability of the system cannot be definitively confirmed yet.

Beyond the simple ratio of orbital periods, the stability of the system could be affected by the interplay of von Zeipel-Lidov-Kozai oscillations in the triple system with the presence and dynamics of circumstellar disks. In order for these oscillations to take place, the misalignment angle between the two orbital planes must be $\Delta\theta \gtrsim40\degr$. One of the two possible deprojection solutions has a modest misalignment angle ($\Delta\theta\approx15\degr$), lower than the threshold for the von Zeipel-Lidov-Kozai mechanism to be triggered. This solution is broadly consistent with the statement that triple systems are generally mildly misaligned when the outer orbit is wider than $\approx50$\,au \citep{tokovinin17}. We also note that in this situation, the ratio of $\sim25-100$ between the inner and outer orbital periods favors a low-to-moderate eccentricity for the outer \ggt\,Aa--Ab system \citep{sterzik02}, which is independently more plausible to account for the inner radius of the circumtriple ring \citep{toci24}. In the alternative deprojection solution, however, von Zeipel-Lidov-Kozai mechanism would be triggered and both the misalignment angles and the orbital eccentricities should go through large oscillations. Given an orbital period of a few centuries and a moderate eccentricity for the \ggt\,Aa-Ab orbit \citep{kohler11, toci24}, the timescale of these oscillations should be $\lesssim10^5$\,yr \citep{hamers21}, significantly shorter than the age of the system but still much longer than the current orbital coverage. 

Finally, we note that we have so far mostly considered \uxt\ as a triple system containing the A and B components. As noted by \cite{zapata20} and \cite{menard20}, the morphology of extended spiral arms in the \uxt\,A disk and the presence of circumstellar dust and gas associated with \uxt\,C point to the strong dynamical influence of the latter. However, considering that \uxt\,B system is an order of magnitude more massive than \uxt\,C \citep{kraus09}, it is possible that the close binary could also have a significant dynamical influence on the \uxt\,A disk. Conversely, it is also possible that \uxt\,C will alter the orbit of \uxt\,B in case a reasonably close fly-by. The relative astrometry between the two components is not currently known sufficiently well to explore this possibility at this stage, however.

%--------------------------------------------------------------------
\section{Conclusion}

We have presented new adaptive optics and interferometric observations of the close binary systems \ggt\,Ab and \uxt\,B, which are both part of higher order multiple pre-main sequence systems. In both cases, the observations provide sufficient orbital coverage ($\approx60-100\%$) to tightly constrain the Keplerian orbits, with orbital periods of $\approx8$ and 32\,yr for \ggt\,Ab and \uxt\,B, respectively. 

Deriving dynamical masses for both systems, we find that \ggt\,Aa and the combined Ab subsystem have similar masses, $\approx0.6-0.8\,M_\odot$, consistent with the known spectral types of the two components and the near-infrared flux ratio of the Ab subsystem. In the case of \uxt\,B, we find its dynamical mass of 1.5\,$M_\odot$ to be somewhat higher than expected but could be reconciled if the two components are of equal mass and of slightly earlier spectral type than reported in the literature. 

The orbital solution we obtain for both systems are eccentric, but markedly more so for \uxt\,B ($e\approx0.87$). Along with the lack of outer reservoir that could feed the system, this high eccentricity could explain the lack of circumstellar dust in the subsystem, as stable orbits are confined to $\lesssim1$\,au or $\gtrsim50$\,au. 

While the orbit of the outer components of the system are moderately constrained (\ggt\,Aa--Ab) or completely unconstrained (\uxt\,A--B), we find it likely that the triple systems are stable on the long term and, thus, that they could contribute to the growing population of exoplanets in multiple systems observed in the field. The most important caveats to this assessment are in the \ggt\ system, where 1) ongoing accretion could further alter the orbits, and 2) one of the possible deprojection solution implies a near-polar configuration for the triple system in which large von Zeipel-Lidov-Kozai oscillations would play a major role. 

In both systems, new, dedicated hydrodynamical simulations are warranted to further explore the dynamics of both the stars and the gas+dust components in both systems. Planet formation in multiple system is an increasingly central topic is modern astronomy and these systems not only represent arguably typical systems but their properties are now known with high precision. Establishing with precision the orbital arrangement of young stars with discs will be pivotal to understand the long term evolution of discs in multiple stellar systems. This aspect is crucial to eventually pinpoint the locations at which planet(esimals) could form in such multiples.

\begin{acknowledgements}
We are grateful to an anonymous referee whose comments helped imprve this manuscript. This project has received funding from the European Research Council (ERC) under the European Union's Horizon Europe research and innovation program (grant agreement No. 101053020, project Dust2Planets; grant agreement No. 101042275, project Stellar-MADE). This research has made use of the Keck Observatory Archive (KOA), which is operated by the W. M. Keck Observatory and the NASA Exoplanet Science Institute (NExScI), under contract with the National Aeronautics and Space Administration. Some of the data presented in this paper were obtained from the Mikulski Archive for Space Telescopes (MAST) at the Space Telescope Science Institute. Some of the data presented in this paper were obtained as part of ESO programs 096.C-0241, 097.C-0865, 198.C-0209, and 0100.C-0452, and retrieved from the ESO archive. M. Langlois acknowledge funding from the French National research agency (grant DDISK, ANR-21-CE31-0015-01).
\end{acknowledgements}


\begin{appendix} %First appendix
\section{Cornerplots for the orbital fits}
\label{sec:appendix}

\begin{figure}
   \centering
   \includegraphics[width=\columnwidth]{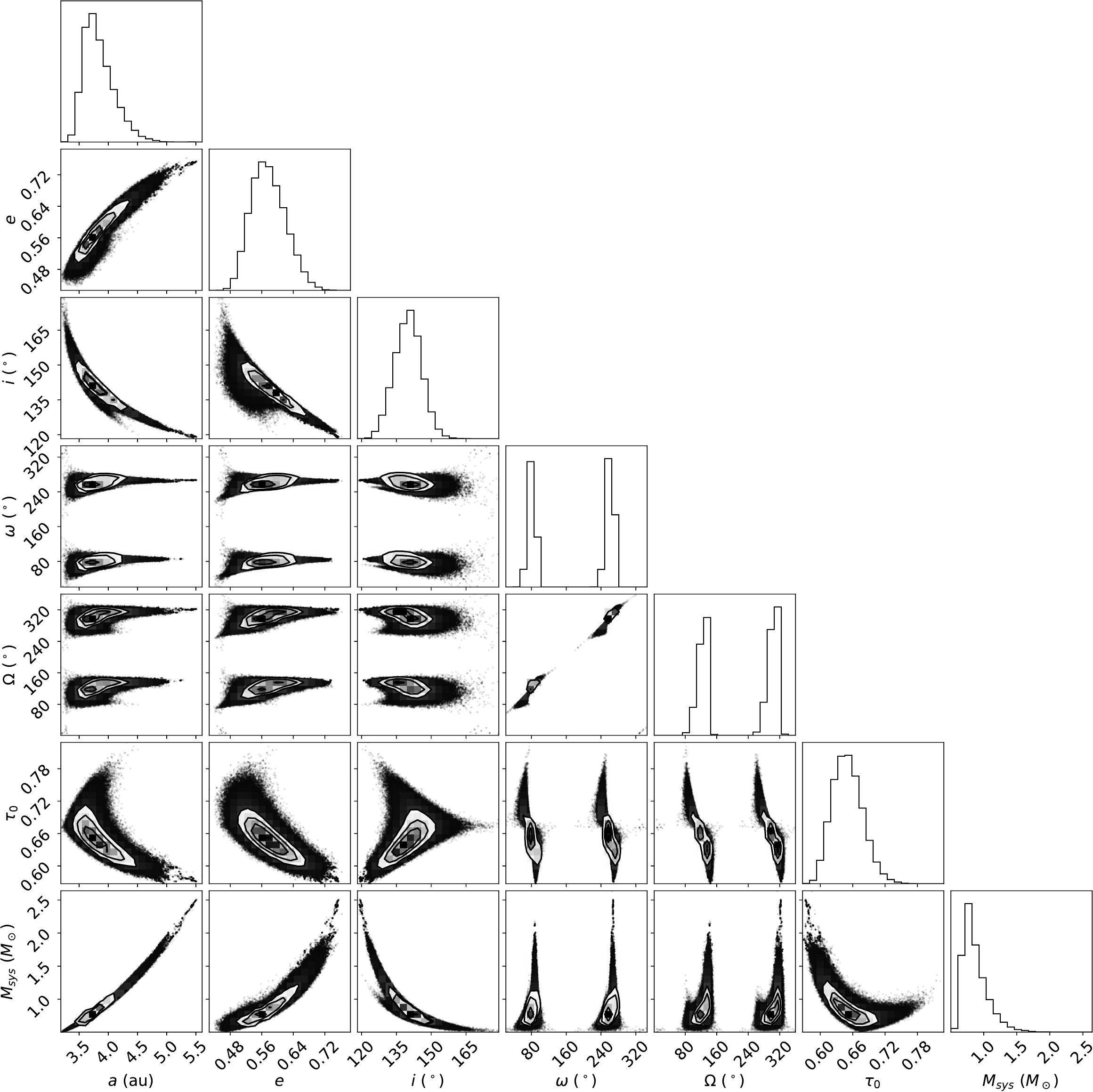}
   \caption{MCMC cornerplot resulting from the orbital fit to the \ggt\,Ab orbit. While {\tt orbitize!} converged on the solution with $i > 90\degr$, we adopt the equivalent supplementary angle, which match the convention of the circumtriple ring.}
   \label{fig:gg_params}
\end{figure}

\begin{figure}
   \centering
   \includegraphics[width=\columnwidth]{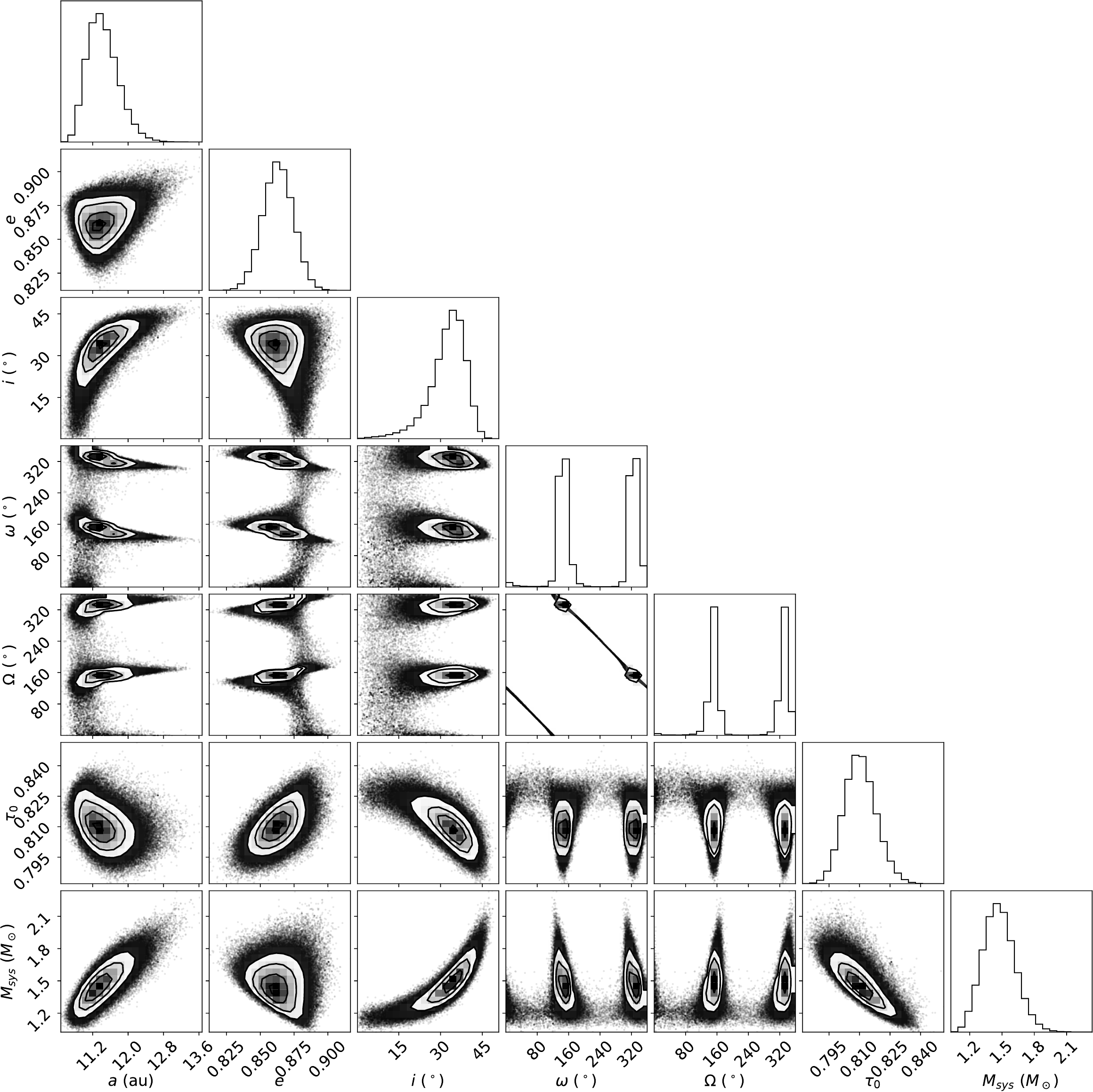}
   \caption{MCMC cornerplot resulting from the orbital fit to the \uxt\,B orbit.}
   \label{fig:ux_params}
\end{figure}

In Figures\,\ref{fig:gg_params} and \ref{fig:ux_params}, we present the MCMC cornerplot resulting from the orbital fits to the \ggt\,Ab and \uxt\,B systems, respectively. The 180\degr\ ambiguity on $\omega$ and $\Omega$ is inherent to visual orbits and the two families are collapsed when computing the confidence intervals on all orbital parameters. We also note that the line of nodes is not defined for $i=0$\degr, which explains the wide spread of solutions for $\omega$ and $\Omega$ at the lowest inclinations for the \uxt\,B system.

\section{Orbits of pre-main sequence visual binaries form the literature}
\label{sec:litt_orbits}

Here we compile the visual pre-main sequence binaries with published orbital solutions. When multiple solutions have been published, we adopted the most recent solution. Only systems with orbital periods $P<50$\,yr are considered here to ensure good orbital coverage, commensurable with that we obtained for \ggt\,Ab and \uxt\,B. We excluded HBC\,351 from the sample as it is most likely to be a member of the 120\,Myr-old Plieades open cluster \citep{luhman09}.

\begin{table*}
\caption{Orbital solutions for visual pre-main sequence binaries. Systems marked with a $\dagger$ symbols are members of the Upper Scorpius or TW Hydra association, whose age ($\sim10$--20\,Myr) is such that the absence of a disk may no longer be a consequence of multiplicity. }
\label{tab:litt_orbits}
\centering
\begin{tabular}{cccccc}
\hline
System & $P$ (yr) & $e$ & $a$ (au) & Disk? & Reference\\
\hline
DF\,Tau & 43.7$\pm$3.0 & 0.29$\pm$0.07 & 13.8$\pm$1.0 & Y & \cite{schaefer14, sullivan22} \\
DI\,Tau & 35.1$^{+36.0}_{-2.9}$ & 0.73$^{+0.13}_{-0.03}$ & 10.6$^{+5.5}_{-0.2}$ & Y & \cite{tang23, sullivan22} \\
FF\,Tau & 14.77$\pm$0.18 & 0.634$\pm$0.004 & 6.25$\pm$0.08 & N & \cite{rizzuto19, sullivan22} \\
GSC\,6209 & 5.52$\pm$0.03 & 0.21$\pm$0.01 & 4.1$\pm$ 0.4 & N$^\dagger$ & \cite{rizzuto16} \\
GSC\,6794 & 13.28$\pm$0.08 & 0.262$\pm$0.004 & 8.4$\pm$0.8 & N$^\dagger$ & \cite{rizzuto16} \\
HP\,Tau\,G3 & 27.33$^{+1.35}_{-1.19}$ & 0.521$^{+0.009}_{-0.008}$ & 9.07$^{+0.15}_{-0.12}$ & N & \cite{rizzuto19, kraus12} \\
Hubble\,4 & 9.287$\pm$0.004 & 0.680$\pm$0.001 & 5.53$\pm$0.02 & N & \cite{rizzuto19, sullivan22} \\
J160517-202420 & 5.88$\pm$0.01 & 0.907$\pm$0.004 & 3.2$\pm$0.3 & N$^\dagger$ & \cite{rizzuto16} \\
NTTS\,045251 & 6.91$\pm$0.03 & 0.46$\pm$0.02 & 4.8$\pm$0.3 & N & \cite{steffen01} \\
ROXs\,47\,A & 8.23$\pm$0.12 & 0.82$\pm$0.01 & 4.7$\pm$0.5 & Y & \cite{rizzuto16} \\
RX\,J1550.0 & 8.74$\pm$0.04 & 0.488$\pm$0.001 & 6.8$\pm$0.7 & N$^\dagger$ & \cite{rizzuto16} \\
RX\,J1601.9 & 8.17$\pm$0.06 & 0.36$\pm$0.01 & 5.3$\pm$0.5 & N$^\dagger$ & \cite{rizzuto16} \\
ScoPMS\,17 & 14.98$\pm$0.13 & 0.880$\pm$0.002 & 6.9$\pm$0.7 & N$^\dagger$ & \cite{rizzuto16} \\
T\,Tau\,S & 27.2$\pm$ 0.7 & 0.55$\pm$0.03 & 12.7$\pm$0.1 & Y & \cite{schaefer20, white01} \\
TWA\,5 & 5.94$\pm$0.09 & 0.78$\pm$0.05 & 2.9$\pm$0.3 & N$^\dagger$ & \cite{konopacky07, riviere13} \\
V773\,Tau\,A--B & 26.2$\pm$1.1 & 0.10$\pm$0.03 & 15.4$\pm$0.5 & N & \cite{boden12, duchene03} \\
V807\,Tau & 12.0$\pm$0.4 & 0.34$\pm$0.06 & 5.0$\pm$0.3 & Y & \cite{galli18, sullivan22} \\
V1000\,Tau & 3.6$\pm$0.3 & 0.43$\pm$0.05 & 2.5$\pm$0.1 & Y & \cite{galli18, sullivan22} \\
V1023\,Tau & 9.30$\pm$0.05 & 0.69$\pm$0.01 & 6.3$\pm$0.1 & N & \cite{galli18, white01} \\
ZZ\,Tau & 46.8$\pm$0.8 & 0.58$\pm$0.02 & 11.6$\pm$0.6 & Y & \cite{belinski22, sullivan22} \\
\hline
\end{tabular}
\end{table*}

\end{appendix}

\end{document}